\documentclass{aa}
\usepackage{graphicx}
\usepackage{threeparttable}
\usepackage{txfonts}

\authorrunning{Zhong et al}
\titlerunning{Open cluster catalogs with Gaia and LAMOST}

\begin{document}

\title{Exploring open cluster properties with Gaia and LAMOST}

\author{Jing Zhong \inst{1},
          Li Chen \inst{1,2},
          Di Wu \inst{3},
          Lu Li \inst{1,2},
          Leya Bai \inst{1,3},
          \and Jinliang Hou \inst{1,2}  }

   \institute{Key Laboratory for Research in Galaxies and Cosmology, Shanghai Astronomical Observatory, Chinese Academy of  Sciences, 80 Nandan Road, Shanghai 200030, China,
   \email{jzhong@shao.ac.cn,chenli@shao.ac.cn}\\
   \and
   School of Astronomy and Space Science, University of Chinese Academy of Sciences, No. 19A, Yuquan Road, Beijing 100049, China \\
   \and
   Physics and Space Science College, China West Normal University, 1 ShiDa Road, Nanchong 637002, China }
   
   \date{}

 \abstract
  {In Gaia DR2, the unprecedented high-precision level reached in sub-mas for astrometry and mmag for photometry. Using cluster members identified with these astrometry and photometry in Gaia DR2, we can obtain a reliable determination of cluster properties. However, because of the shortcoming of Gaia spectroscopic observation in dealing with densely crowded cluster region, the number of radial velocity and metallicity for cluster member stars from Gaia DR2 is still lacking. It is necessary to combine the Gaia data with the data from large spectroscopic surveys, such as LAMOST, APOGEE, GALAH, Gaia-ESO, etc.}
   {In this study, we aim to improve the cluster properties by combining the LAMOST spectra. In particular, we provide the list of cluster members with spectroscopic parameters as an add-value catalog in LAMOST DR5, which can be used to perform detailed study for a better understanding on the stellar properties, by using their spectra and fundamental properties from the host cluster.}
   {We cross-matched the spectroscopic catalog in LAMOST DR5 with the identified cluster members in \citet[hereafter CG18]{2018A&A...618A..93C}. We then used members with spectroscopic parameters to derive statistical properties of open clusters.}
   {We obtained a list of 8811 members with spectroscopic parameters and a catalog of 295 cluster properties. In addition, we study the radial and vertical metallicity gradient and age-metallicity relation with the compiled open clusters as tracers, finding slopes of -0.053$\pm$0.004 dex kpc$^{-1}$, -0.252$\pm$0.039 dex kpc$^{-1}$ and 0.022$\pm$0.008 dex Gyr$^{-1}$, respectively. Both slopes of metallicity distribution relation for young clusters (0.1 Gyr $<$ Age $<$ 2 Gyr) and the age-metallicity relation for clusters within 6 Gyr are consistent with literature results. In order to fully study the chemical evolution history in the disk, more spectroscopic observations for old and distant open clusters are needed for further investigation.}
  {}
  
 \maketitle 
%___________________________

\section{Introduction}
\label{intro}

Open clusters are ideal tracers to study the stellar population, the Galactic environment, and the formation and evolution of Galactic disk. Open clusters have large age and distance spans and can be relatively accurately dated; the spatial distribution and kinematic properties of OCs provide critical constraints on the overall structure and dynamical evolution of the Galactic disk. Meanwhile, their [M/H] values serve as excellent tracers of the abundance gradient along the Galactic disk, as well as many other important disk properties, such as the age-metallicity relation (AMR), abundance gradient evolution, etc \citep{1979ApJS...39..135J,1995ARA&A..33..381F,1993A&A...267...75F,1998MNRAS.296.1045C,2002AJ....124.2693F,2008A&A...480...79B,2008A&A...488..943S,2009A&A...494...95M,2010AJ....139.1942F,2011A&A...535A..30C,2016MNRAS.463.4366R}. 

Most open clusters are located on the galactic disk. Up to now, about 3000 star clusters have been cataloged \citep{2002A&A...389..871D, 2013A&A...558A..53K} including about 2700 open clusters, most of which were located within 2-3~kpc of the Sun. However, limited by the precision of astrometric data, for many of those cataloged open clusters the reliability of member-selection and thereby the derived fundamental parameters had remained being uncertain. The European Space Agency (ESA) mission {\it Gaia} ({\it https://www.cosmos.esa.int/gaia}) implemented an all-sky survey, which has released its Data Release 2  \citep[Gaia-DR2;][]{2018A&A...616A...1G} providing precise five astrometric parameters (positions, parallaxes, and proper motions) and three band photometry ($G$, $G_{BP}$ and $G_{RP}$ magnitude) for more than one billion stars \citep{2018A&A...616A...2L}. Using the astrometry and photometry of Gaia DR2, cluster members and fundamental parameters of open clusters have been determined with high level of reliability \citep{2018A&A...618A..93C, 2018A&A...619A.155S, 2019A&A...623A.108B, 2019AstL...45..208B}. Furthermore, the unprecedented high precision astrometry in Gaia DR2 is also can be used to discover new open clusters in the solar neighborhood \citep{2018A&A...618A..59C,2019A&A...624A.126C, 2019MNRAS.483.5508F}, as well as the extended substructures in the outskirts of open clusters \citep{2019A&A...624A..34Z,2019A&A...621L...2R,2019A&A...621L...3M}.

Although Gaia DR2 provide accurate radial velocities for about 7.2 million FGK stars, it is incomplete in terms of radial velocities, providing them only for the brightest stars. The observational mode of slitless spectroscopy of Gaia made it hard to observe densely crowded regions, since multiple overlapping spectra would be noisy and make the deblending process very difficult \citep{2018A&A...616A...5C}. Using the weighted mean radial velocity based on Gaia DR2, \citet[ hereafter SC18]{2018A&A...619A.155S} reported the 6D phase space information of 861 star clusters. However, about 50\% clusters only have less than 3 member stars with radial velocity available.    

As an ambitious spectroscopic survey project, the Large sky Area Multi-Object fiber Spectroscopic Telescope \citep[LAMOST,][]{Cui2012,Zhao2012,Luo2012} provided about 9 million spectra with radial velocities in its fifth data-release (DR5), including 5.3 million spectra with stellar atmospheric parameters (effective temperature, surface gravity and metallicity) derived by LAMOST Stellar Parameter Pipeline (LASP). In order to study the precision and uncertainties of atmospheric parameters in LAMOST, \citet{2015RAA....15.1095L} performed the comparison for 1812 common targets between LAMOST and SDSS DR9, and provided the measurement offsets and errors as: -91$\pm$111 K in effective temperature (T$_{\rm eff}$), 0.16 $\pm$ 0.22 dex in surface gravity (Log$g$), 0.04 $\pm$ 0.15 dex in metallicity ([Fe/H]) and -7.2 $\pm$ 6.6 km s$^{-1}$ in radial velocity (RV). Since most of observations in LAMOST were focus on the Galactic plane, we expect to obtain the full 3D velocities information for members of hundreds open clusters in the Galactic Anti-Center.

In this paper, our main goals are to derive the properties of open clusters based on Gaia DR2 and LAMOST data, and to provide a catalog of spectroscopic parameters of cluster members. In section~\ref{s}, we describe how we derived the cluster properties, including radial velocities, metallicities, ages, and 6D kinematic and orbital parameters.  Using the sample of 295 open clusters, we investigate their statistic properties, and study the radial metallicity gradient and the age-metallicity relation in section~\ref{p}. A brief description of the catalogs of the clusters and their member stars are presented in section~\ref{cat}.   

\begin{figure}
   \centering
   \includegraphics[angle=0,scale=1.1]{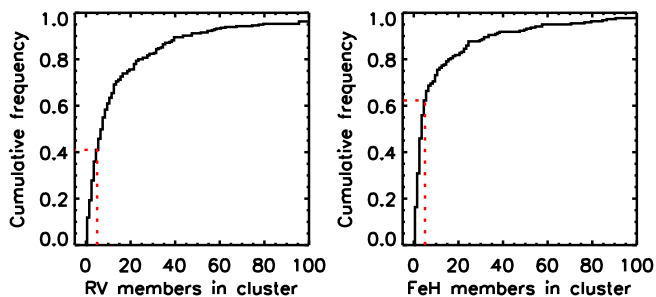}
  % \begin{minipage}[]{85mm}
   \caption{Left panel: cumulative number of RV members in 295 open clusters. About 59\% clusters have RV members greater than 5. Right panel: cumulative number of [Fe/H] members in 220 open clusters.  About 38\% clusters have [Fe/H] members greater than 5. }
%\end{minipage}
   \label{his_mem}
\end{figure}

\section{The sample}
\label{s}

\subsection{Members and cluster parameters}
We choose the open cluster catalog and their member stars of CG18 as our starting sample. In this catalog, a list of members and astrometric parameters for 1229 clusters were provided, including 60 newly discovered clusters. 

In order to identify cluster members, CG18 applied a code called UPMASK \citep{2014A&A...561A..57K} to determine the membership probability of stars located on the cluster field. Based on the unprecedentedly precision Gaia astrometric solution ($\mu_{\alpha},~ \mu_{\delta}, ~\varpi$ ), those cluster members were believed to be well identified with highly reliability. A total of 401,448 stars were provided by CG18, with membership probabilities ranging from 0.1 to 1.

Once cluster members were obtained, the mean astrometric parameters of clusters such as proper motions and distance were derived. In CG18, the cluster distances were estimated from the Gaia DR2 parallaxes, while the fractional uncertainties $\sigma_{\langle \varpi \rangle}$ / $\langle \varpi \rangle $ for 84\% clusters are below 5\%. 

\subsection{Radial velocities}
\label{radial velocity}

%  RVcompiled
\begin{figure*}
   \centering
   \includegraphics[angle=0,scale=1.1]{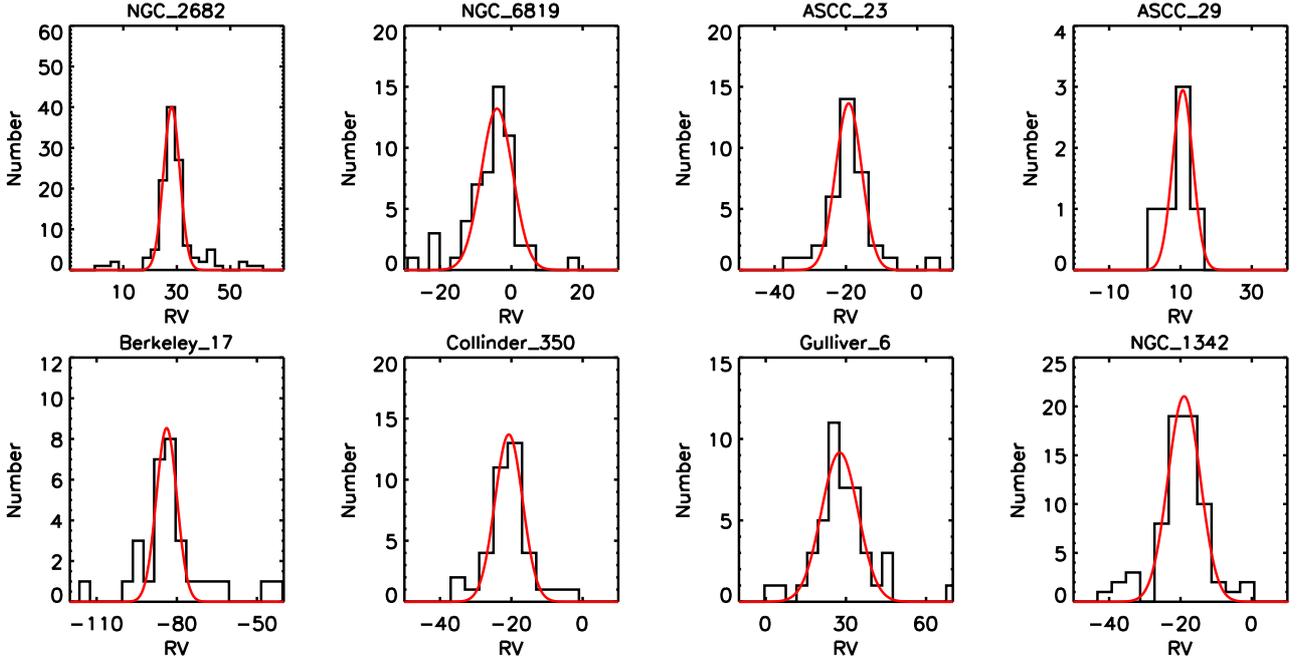}
  % \begin{minipage}[]{85mm}
   \caption{Radial velocity distribution and fitting profile for each open cluster. The complete figures of fitting result are available alongside the article.}
%\end{minipage}
   \label{rv}
\end{figure*}

Using the member stars provided by CG18, we perform the cross-matching process with the LAMOST DR5 by a radius of 3". A number of 8811 stars were identified as having the LAMOST spectra, while 3935 of them have atmospheric parameters with high signal-to-noise ratio (SNR in $g$ band $\geq$ 15 for A,F,G type stars and SNR in $g$ band $\geq$ 6 for K-type stars ) . The uncertainty of RV provided by LAMOST is about 5 km s$^{-1}$ \citep{2015MNRAS.448..822X,2015RAA....15.1095L}.

In order to derive the average radial velocity for each open cluster, we only select stars whose membership probabilities greater than 0.5 and have RV parameter available in in LAMOST DR5. A total of 6017 stars in 295 cluster were left for average RV calculation. The left panel in Figure~\ref{his_mem} shows the cumulative number distribution of RV members in 295 open clusters. In our cluster sample, the number of RV members of 174 cluster (59 \%) is greater than 5, which indicate the higher reliability of derived RV parameters for these clusters. 

It is not suitable to simply use the mean RV of members as the overall RV of an open cluster. This is because the mean RV is easy to be contaminated by misidentified member stars (in fact they are field stars with different RVs) or member stars with large RV measurement uncertainties (e.g., stars of  early type  or late type, or stars with low SNR). The mean RV of members will have large uncertainties and lead to unpredictable offsets, especially for clusters with only a few RV members. 

To solve this problem and derive a reliable average RV for open clusters, we carefully check the RV distribution histogram of each open cluster and for those with sufficient RV data we use a Gaussian profile to fit the RV distribution of member stars. Outliers will be excluded in the Gaussian fitting process. For each cluster, the $\mu$ and $\sigma$ of Gaussian function are used as the average RV and corresponding uncertainty. Figure~\ref{rv} shows a few examples of the RV fitting results. In our sample, clusters which have the average RV estimation derived by the Gaussian fitting process are marked as the high quality samples with the RV\_flag labeled as 'Y' in the catalog (See Table~\ref{cat_ocs}).  On the other hand, for clusters which were suffered with small RV members or have large dispersion in RV distribution, we simply provide mean RVs and standard deviations as their overall RVs and uncertainties, respectively.     

\subsection{Metallicities}

%  FeH
\begin{figure*}
   \centering
   \includegraphics[angle=0,scale=1.1]{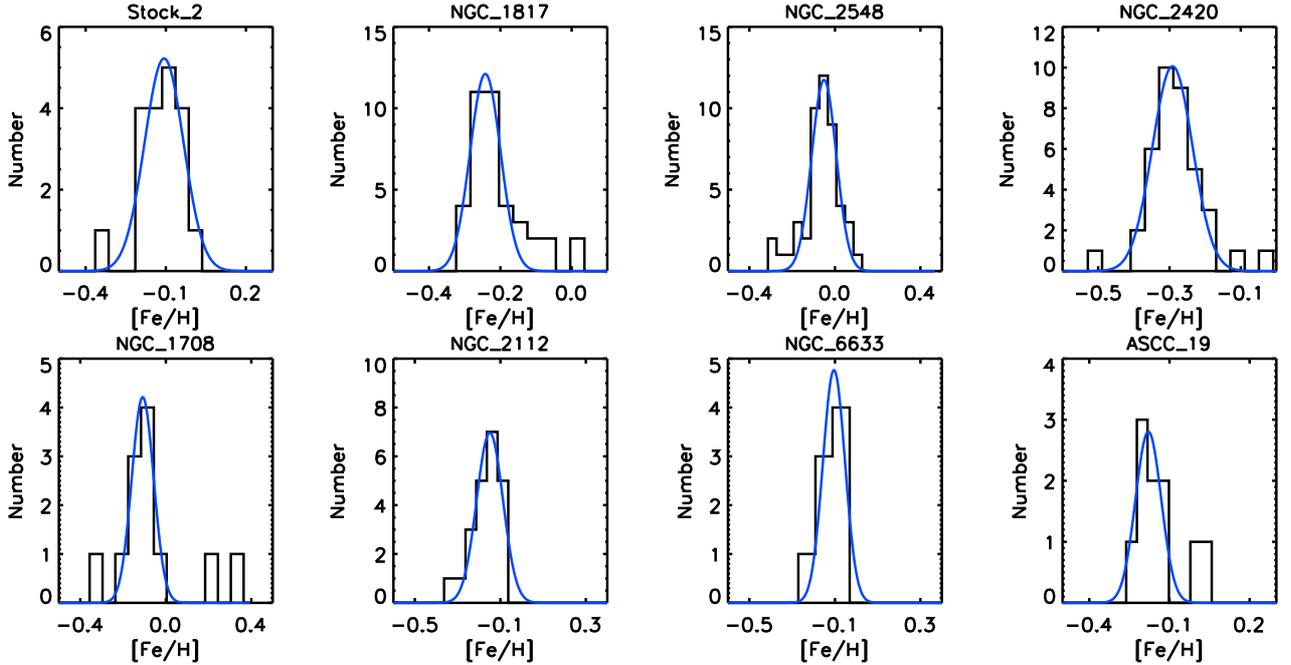}
  % \begin{minipage}[]{85mm}
   \caption{[Fe/H] metallicity distribution and fitting profile for each open cluster. The complete figures of fitting result are available alongside the article.}
%\end{minipage}
   \label{feh}
\end{figure*}

\begin{figure*}
   \centering
   \includegraphics[angle=0,scale=1.1]{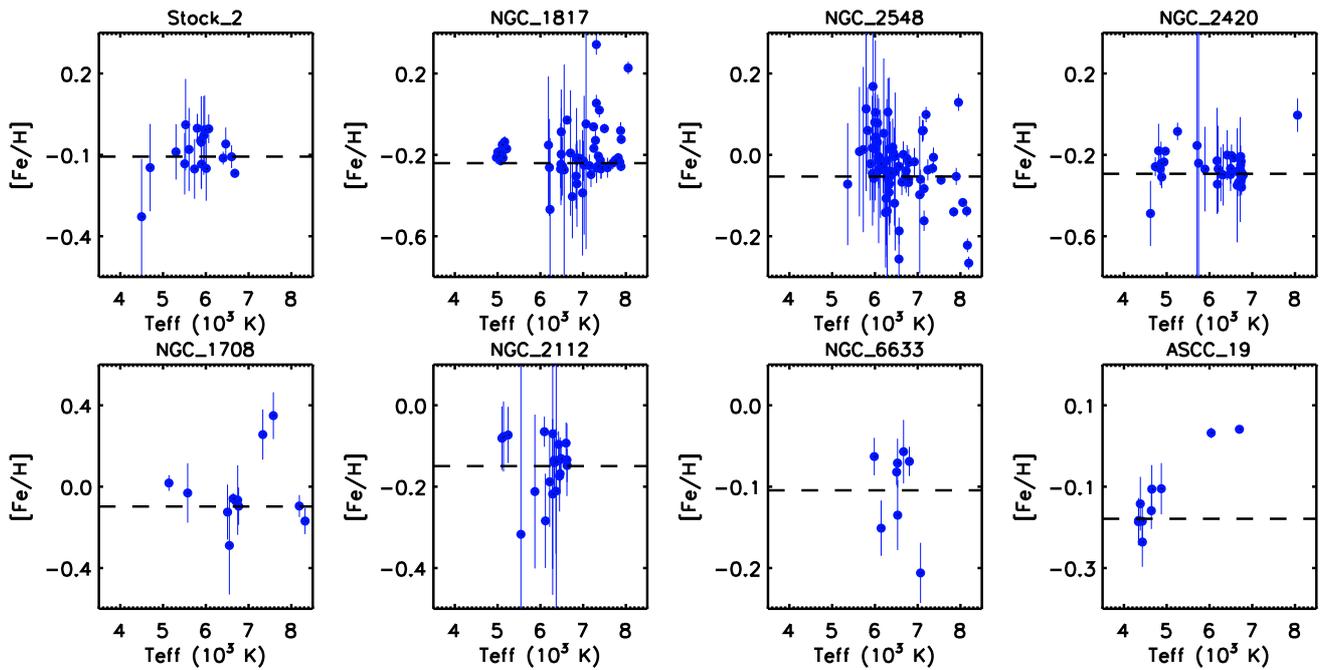}
  % \begin{minipage}[]{85mm}
   \caption{[Fe/H] metallicity distribution as a function of temperature (Teff) for each open cluster. Dashed line represent the overall [Fe/H] metallicity derived by Gaussian fitting in Figure~\ref{feh}.}
%\end{minipage}
   \label{feh_teff}
\end{figure*}

\begin{figure*}
   \centering
   \includegraphics[angle=0,scale=1.1]{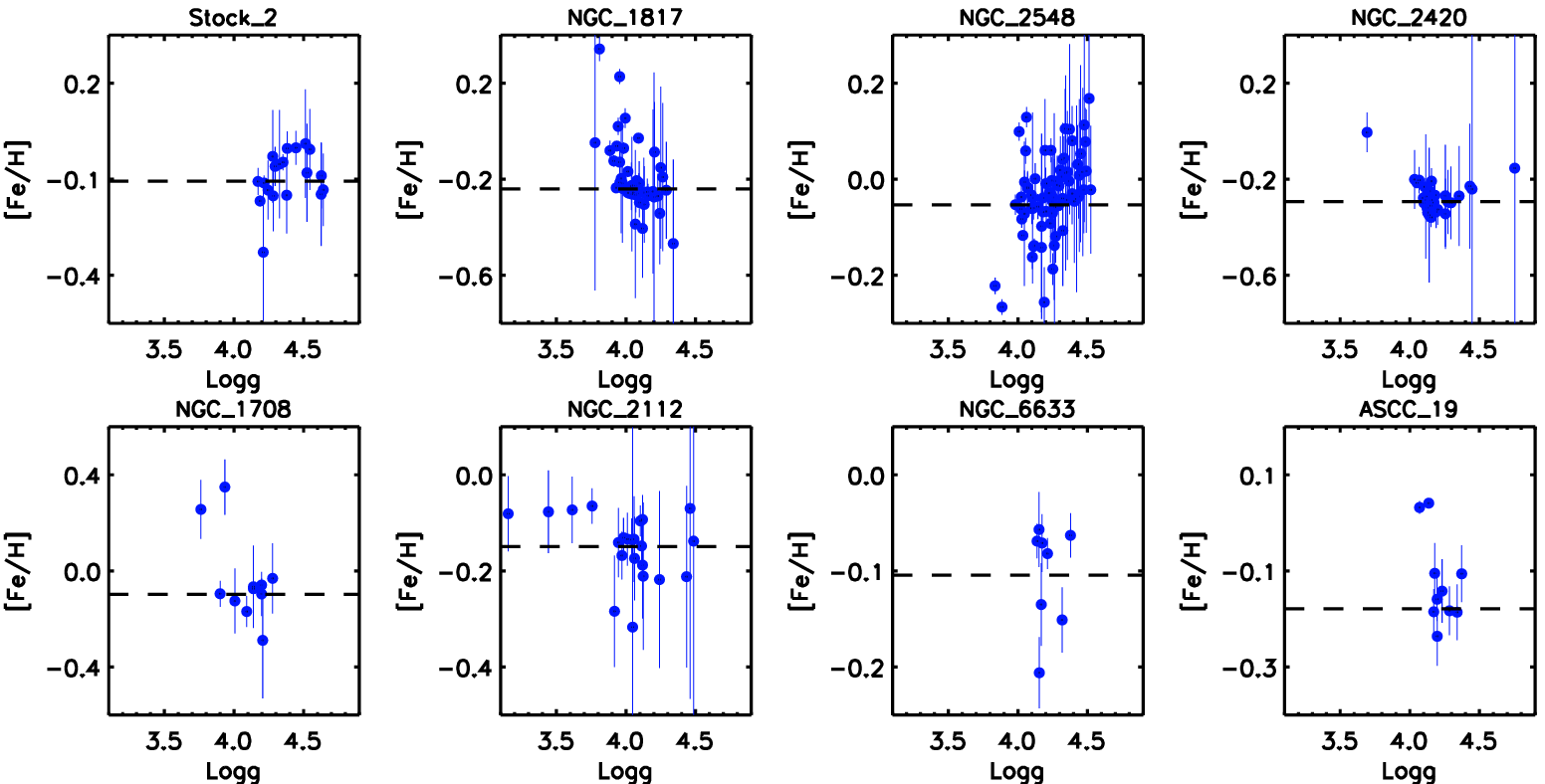}
  % \begin{minipage}[]{85mm}
   \caption{[Fe/H] metallicity distribution as a function of surface gravity (Logg) for each open cluster. Dashed line represent the overall [Fe/H] metallicity derived by Gaussian fitting in Figure~\ref{feh}.}
%\end{minipage}
   \label{feh_logg}
\end{figure*}

The fifth data release of LAMOST (DR5) provides a stellar parameters catalog including 5.3 million spectra \citep{2015RAA....15.1095L}. Following the determination process of the overall RV of open clusters, we first cross-match cluster members of CG18 with the stellar parameters catalog in LAMOST. Then, we select stars with membership probabilities greater than 0.5 and have [Fe/H] measurements available, 3024 stars in 220 clusters were selected for metallicity estimation.
 
Using members with [Fe/H] measurement, we plot the metallicity distribution histogram and perform the Gaussian fitting for each open cluster. As we have done in the RV estimation, outliers which have very different metallicity values were excluded by visual inspection. A few examples of the fitting results were presented in Figure~\ref{feh}, while the $\mu$ and $\sigma$ of Gaussian function are used as the average metallicity and corresponding uncertainty respectively. For the rest of open clusters, whose metallicity distribution can not be fitted by the Gaussian function, their overall metallicities and uncertainties are set as the mean [Fe/H] and standard deviations respectively. 

In order to further understand the internal consistency and  parameter independence of [Fe/H] metallicity of LAMOST DR5, we study the [Fe/H] distribution as a function of Teff and Logg. Using the same clusters in Figure~\ref{feh} as examples, Figure~\ref{feh_teff} and Figure~\ref{feh_logg}  show [Fe/H]  Vs. Teff and [Fe/H] Vs. Logg  results, respectively.  Although there are a few outliers or stars with large [Fe/H] measurement errors, there is no apparent degeneracy between [Fe/H] and other parameters, and the fitting results (dashed line) properly represent the overall metallicity of these clusters. 

%  ISO
\begin{figure*}
   \centering
   \includegraphics[angle=0,scale=1.1]{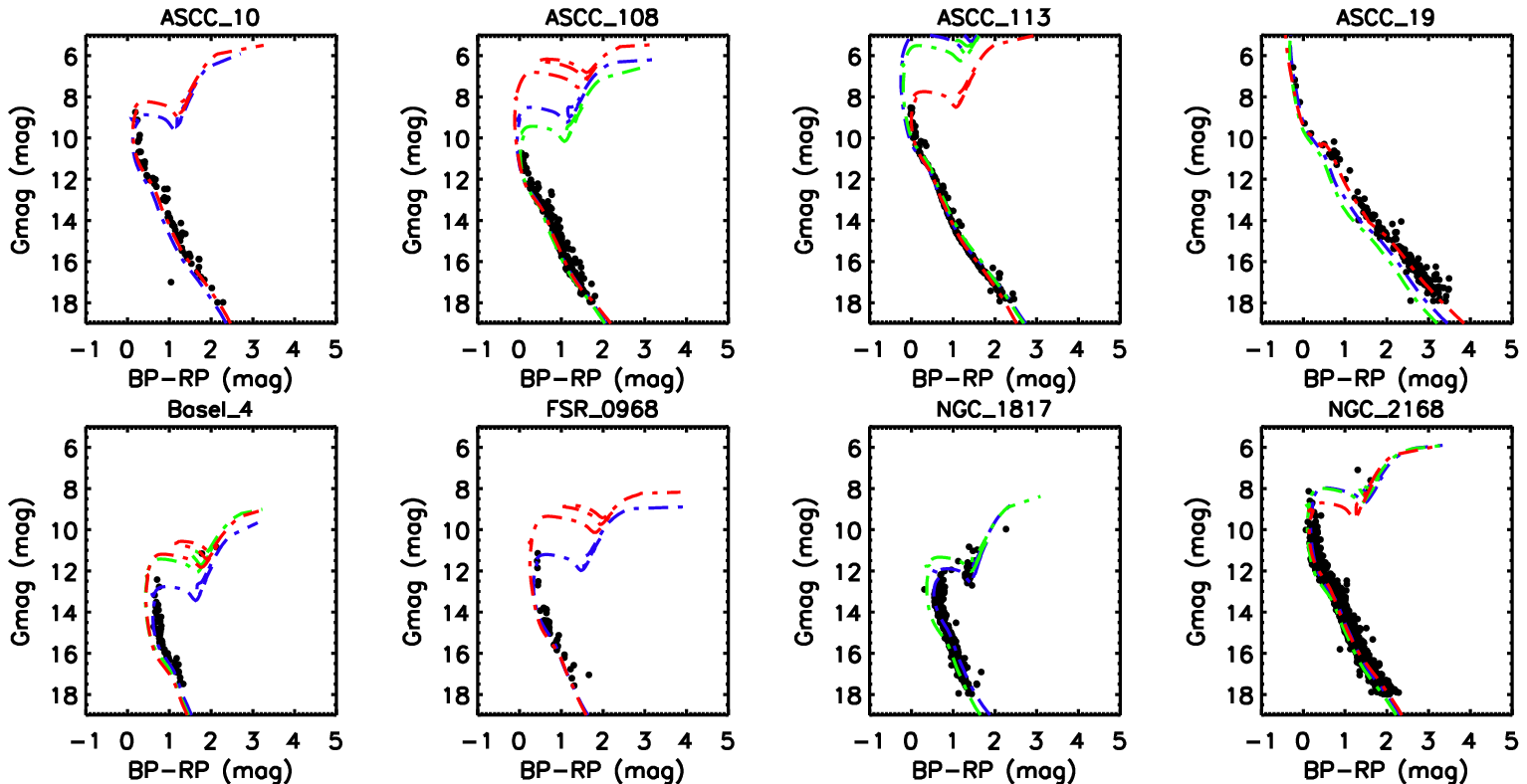}
  % \begin{minipage}[]{85mm}
   \caption{Examples of members distribution in color-magnitude diagram. Colors are represent isochrone parameters which provided by different literatures: \citet{2002A&A...389..871D} in green, \citet{2012A&A...543A.156K} in blue and \citet{2019A&A...623A.108B} in red. The complete figures of fitting result are available alongside the article.}
%\end{minipage}
   \label{iso}
\end{figure*}

\begin{figure*}
   \centering
   \includegraphics[angle=0,scale=1.1]{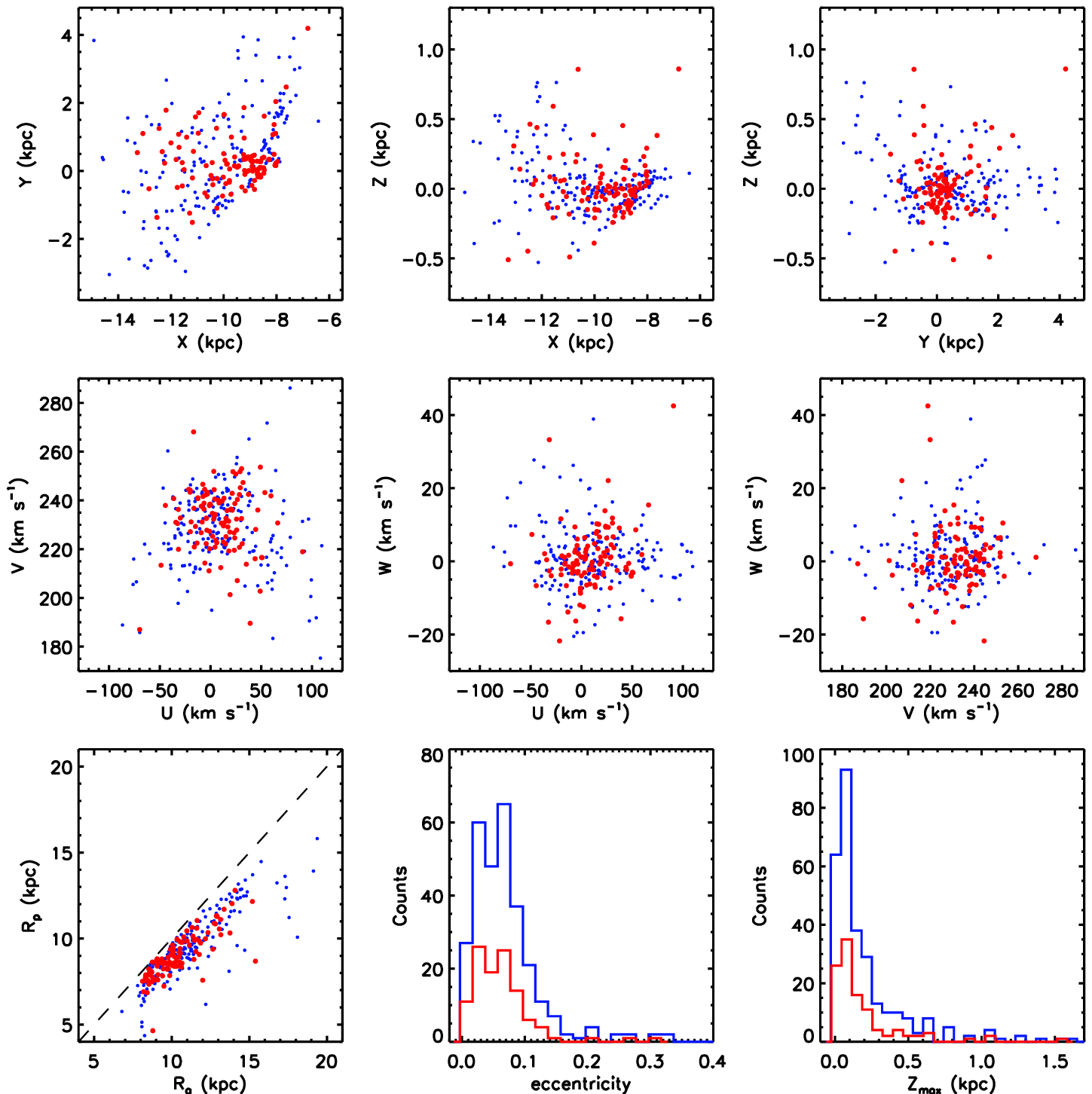}
   \caption{Distribution of derived spatial and kinematic parameters. Blue dots are 295 open clusters with radial velocity estimations. Red dots are 109 open clusters which have radial velocity measurements with high quality. The distribution of clusters illustrate that they are located on the Galactic plane and have kinematics typical of the thin disk.  }
   \label{parm}
\end{figure*}

\subsection{Ages}
\label{age}
In order to provide the age parameter of our sample clusters, we have utilized literature results from  \citet{2002A&A...389..871D,2012A&A...543A.156K,2019A&A...623A.108B} to perform the isochrone fitting and visually determine best fitting result of the age, distance and reddening parameters. Since membership probabilities provided by CG18 are more reliable than previous works, member stars used for isochrone fitting were come from CG18 with probability greater than 0.5. We only provide literature parameters whose isochrone is consistent with the distribution of cluster members in the color-magnitude diagram. In other words, if the age parameter of a cluster in our catalog is zero, that means none of the literature parameters can meet the distribution of cluster members properly. 

Figure~\ref{iso} presents a few examples of the isochrone fitting results. Colors are used to represent three different literature parameters as \citet{2002A&A...389..871D} in green, \citet{2012A&A...543A.156K} in blue and \citet{2019A&A...623A.108B} in red. 

\begin{figure}
   \centering
   \includegraphics[angle=0,scale=1.5]{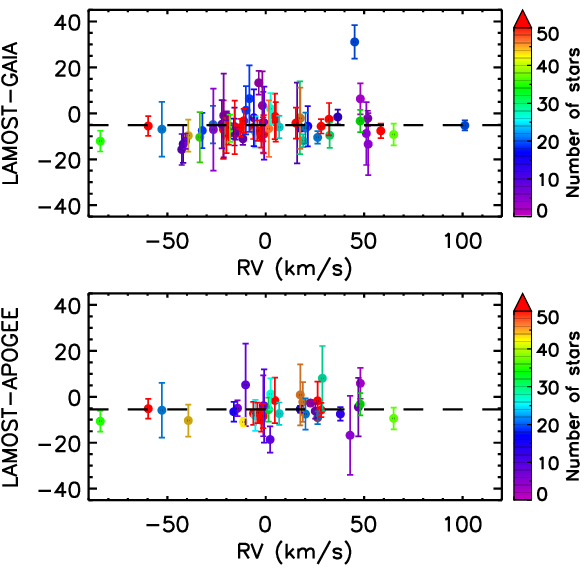}
  % \begin{minipage}[]{85mm}
   \caption{ Upper panel: RV difference for 71 common clusters between SC18 and our catalog.  Bottom panel: RV difference for 36 common clusters between DJ20 and our catalog. The solid circles and their corresponding error bars represent the mean RV and dispersion of each cluster in our catalog, respectively. The color of the data points represents the number of stars used to estimate the average in our catalog. As comparison results of overall RV of open clusters, the average difference for LAMOST-Gaia and LAMOST-APOGEE are -5.1$\pm$6.4 km s$^{-1}$ and -5.5$\pm$5.4 km s$^{-1}$ respectively. }
%\end{minipage}
   \label{cmp_rv}
\end{figure}

\begin{figure}
   \centering
   \includegraphics[angle=0,scale=0.9]{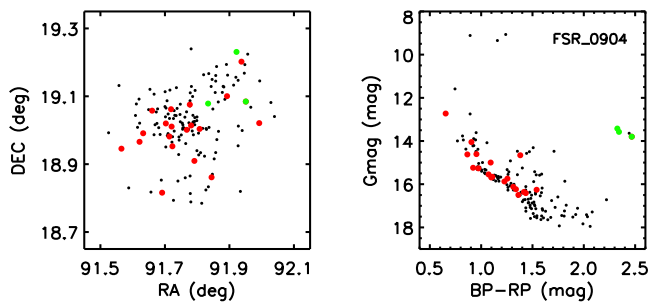}
  % \begin{minipage}[]{85mm}
   \caption{Spatial distribution (left panel) and color-magnitude distribution (right panel) of member stars of FSR\_0904. Black dots are cluster members in CG18. Green and red dots are member stars used for RV estimation in SC18 and our catalog, respectively. It is clear that our RV value of this cluster is more reliable since most of our stars are more likely to be cluster members.}
%\end{minipage}
   \label{fsr0904}
\end{figure}

\begin{figure}
   \centering
   \includegraphics[angle=0,scale=1.1]{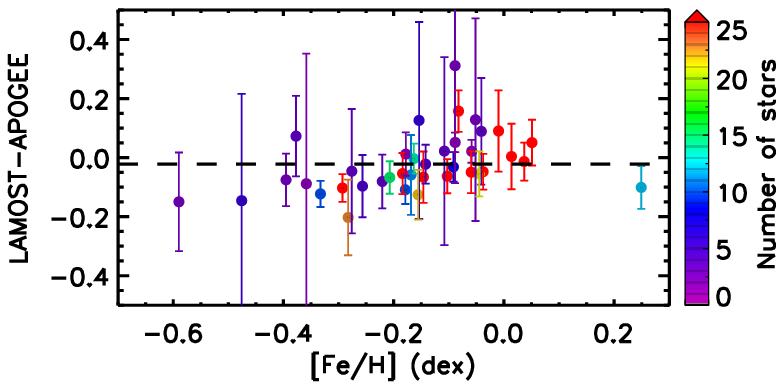}
  % \begin{minipage}[]{85mm}
   \caption{ [Fe/H] difference for 38 common clusters between DJ20 and our catalog. The solid circles and their corresponding error bars represent the mean [Fe/H] and dispersion of each cluster in our catalog, respectively. As an comparison result of overall [Fe/H] of open clusters, the average difference is -0.02$\pm$0.10 dex.}
%\end{minipage}
   \label{cmp_feh}
\end{figure}

\subsection{kinematic parameters}
We calculated the  Galactocentric cartesian coordinates (X, Y, Z) and velocities (U, V, W) of 295 open clusters by using formulas in \citet{1987AJ.....93..864J}. The celestial coordinates, distance and proper motions of each cluster are from CG18, while the radial velocity is determined from the LAMOST DR5 (See section~\ref{radial velocity}). We adopt the solar position and its circular rotation velocity as R$_0$=-8.34 kpc and $\Theta_0$=240 km s$^{-1}$ respectively \citep{2014ApJ...783..130R}. In order to correct for the solar motion in the local standard of rest, we adopt the solar peculiar velocity as (U$_\odot$, V$_\odot$, W$_\odot$)= (11.1, 12.4, 7.25) km s$^{-1}$ \citep{2010MNRAS.403.1829S}.

Based on the astrometry parameters from Gaia DR2 and LAMOST DR5, we further calculated the orbital parameters of 295 open clusters making use of galpy\footnote{http://github.com/jobovy/galpy} \citep{2015ApJS..216...29B}. The orbital parameters are listed in Table~\ref{cat_ocs}, including apogalactic (R$_{\rm ap}$) and perigalactic (R$_{\rm peri}$) distances from the Galactic centre, orbital eccentricity ($e$), and the maximum vertical distance above the Galactic plane (Z$_{\rm max}$).

Figure~\ref{parm} show the distribution of derived spatial and kinematic parameters (blue dots). In particular, we use red color to represent 109 clusters which have radial velocity estimations with high quality  ('RV\_flag' marked 'Y', see section~\ref{radial velocity}). Kinematic parameters,specifically  orbital parameters of these clusters (red dots) are more reliable than others. The Galactocentric spatial distribution  of 295 open clusters in our catalog are shown in the top panels. We find that most of clusters are located on the Galactic anti-center, this is because a large number of LAMOST observational fields are focused on this region. The Galactocentric velocities of open clusters are shown in middle panels. In particular, we exclude 6 open clusters from the velocity and the orbital parameters distribution (bottom panels), since their unreliable radial velocities led to outliers of kinematic parameters. In bottom panels, the distribution of orbital parameters show that most of open clusters have approximate circular motions and small distance to the Galactic plane. Specifically, the kinematic  distribution diagrams clearly illustrate that most of open clusters in our catalog are kinematically typical thin disk.

\subsection{ Comparison to the other works }

To verify the reliability and accuracy of the cluster properties derived by LAMOST DR5, we employed clusters in common between our catalog and other literature catalogs which have high-resolution observation. 

\subsubsection{ Verifying radial velocities }

As we described in Section~\ref{intro}, Gaia DR2 also include accurate radial velocities for 7.2 million stars, which provided by the high-resolution slitless spectrograph (R=11500). SC18 published mean RV for 861 star clusters using spectral results from the Gaia DR2. We use our catalog to crossmatch with SC18 and obtain 218 common clusters. In order to use reliable clusters in SC18 as reference, our comparison only include 83 common clusters which defined as the high quality clusters (see more detail in SC18). In addition, we further exclude 12 common clusters  since their mean RV in our catalog are unreliable (uncertainty greater than 20 km s$^{-1}$). Finally, the number of common clusters used for comparison is 71.

Figure~\ref{cmp_rv} (upper panel) shows the RV difference between SC18 and our catalog for open clusters in common. The average offset of RV is -5.1  km s$^{-1}$ with a scatter of 6.4 km s$^{-1}$. In general, this result shows good agreement with Gaia. The scatter is mainly caused by the RV uncertainties of  LAMOST spectra (R=1800, $\sigma \sim$ 5 km s$^{-1}$), and the number of LAMOST stars in a cluster that used for mean RV estimation (red dots has less scatter than violet dots).

In particular, we note that there is an outlier (blue dot in the upper panel of Figure~\ref{cmp_rv}) with discrepant RV greater than 20 km s$^{-1}$, which named FSR\_0904. After carefully checking the RV data of two catalogs,  we find the number of stars for mean RV estimation is 3 for SC18 and 20 for our catalog. Figure~\ref{fsr0904} shows the spatial distribution and color-magnitude distribution of member stars which were used by two works. At least for this cluster, although the scatter of mean RV in our catalog (7.2 km s$^{-1}$) is greater than in SC18 (2.66 km s$^{-1}$), it is more reliable for the mean RV which provided by our catalog since our stars are mainly distribute on the cluster center and follow the cluster main sequence. 

In addition, we use our catalog and the APOGEE catalog \citep[here after DJ20]{2020AJ....159..199D} to perform the comparison of mean RV and mean [Fe/H] abundance. There are 128 open clusters published by DJ20, including mean RV and mean abundances from the APOGEE DR16. After cross-matching with two catalogs, our sample includes 48 open clusters in common with DJ20. 6 open clusters were further excluded since their 'qual' in DJ20 are flagged  as '0' or 'potentially unreliable'. 

For the comparison of mean RV difference with the APOGEE catalog, 36 common clusters, whose RV uncertainty in our catalog are less than 20 km s$^{-1}$, are plotted in the bottom panel of Figure~\ref{cmp_rv}. The average offset of RV is -5.5 km s$^{-1}$ with a scatter of 5.4 km s$^{-1}$. Similarly as compared with the Gaia result, our mean RV results of clusters are also consistent with the APOGEE catalog, especially for clusters which have more stars to estimate the mean values. 

We note that there are similar RV offsets between our catalog and literature catalogs (SC18 and DJ20), with around -5 km s$^{-1}$. In order to understand the origin and amount of this offset in LAMOST, we perform a general cross-match of stars between LAMOST DR5 and other spectroscopic catalogs (GALAH DR2, APOGEE DR16 and Gaia DR2). Table~\ref{rv_offset} shows the results of RV difference for common stars whose SNR in LAMOST are greater than 10. Here we list the median RV offset, the mean RV offset, standard deviation of RV difference and number of common stars that used for calculation. The similar comparison results of general stars and open clusters show that the RV difference are mainly from the measurement of LAMOST spectra. In addition, we study the RV offset as a function of stellar atmospheric parameters and find that the RV offset is almost a constant all over the parameter space. The result of RV different is also consistent with the conclusion of LAMOST LSP3 parameters analysis \citep{2015MNRAS.448..822X}.

\begin{table}
\caption{Difference of RV for general common stars between LAMOST DR5 and other spectroscopic catalogs.}
\label{rv_offset}
 \centering
\begin{threeparttable}
\begin{tabular}{ccccc}
\hline
 Catalog & Median  & Mean & $\sigma$ & Number\\
 & km s$^{-1}$ & km s$^{-1}$ &   km s$^{-1}$ &  \\
\hline
GALAH~\tnote{1}  &  -4.9  &  -4.8  & 10.6  & 12538 \\
APOGEE~\tnote{2}  &  -4.7  & -4.3   & 9.8 & 96459 \\
Gaia~\tnote{3}    &   -4.9 &  -5.0  & 8.2  & 689838 \\
\hline
\end{tabular}
\begin{tablenotes}
       \footnotesize
       \item[1] \citet{2018MNRAS.478.4513B} 
       \item[2] \citet{2019arXiv191202905A} 
       \item[3] \citet{2018A&A...616A...1G}
\end{tablenotes}
\end{threeparttable}
\end{table}

\begin{table}
\caption{Difference of [Fe/H] for general common stars between LAMOST DR5 and other spectroscopic catalogs.}
\label{feh_offset}
 \centering
\begin{threeparttable}
\begin{tabular}{ccccc}
\hline
 Catalog & Median  & Mean & $\sigma$ & Number\\
 & dex & dex &   dex  &  \\
\hline
GALAH~\tnote{1}  &  0.01  &  0.01  & 0.13  & 11968 \\
APOGEE~\tnote{2}  &  -0.001  & -0.002   & 0.11 & 84355 \\
\hline
\end{tabular}
\begin{tablenotes}
       \footnotesize
       \item[1] \citet{2018MNRAS.478.4513B} 
       \item[2] \citet{2019arXiv191202905A} 
\end{tablenotes}
\end{threeparttable}
\end{table}

\begin{figure*}
   \centering
   \includegraphics[angle=0,scale=1.]{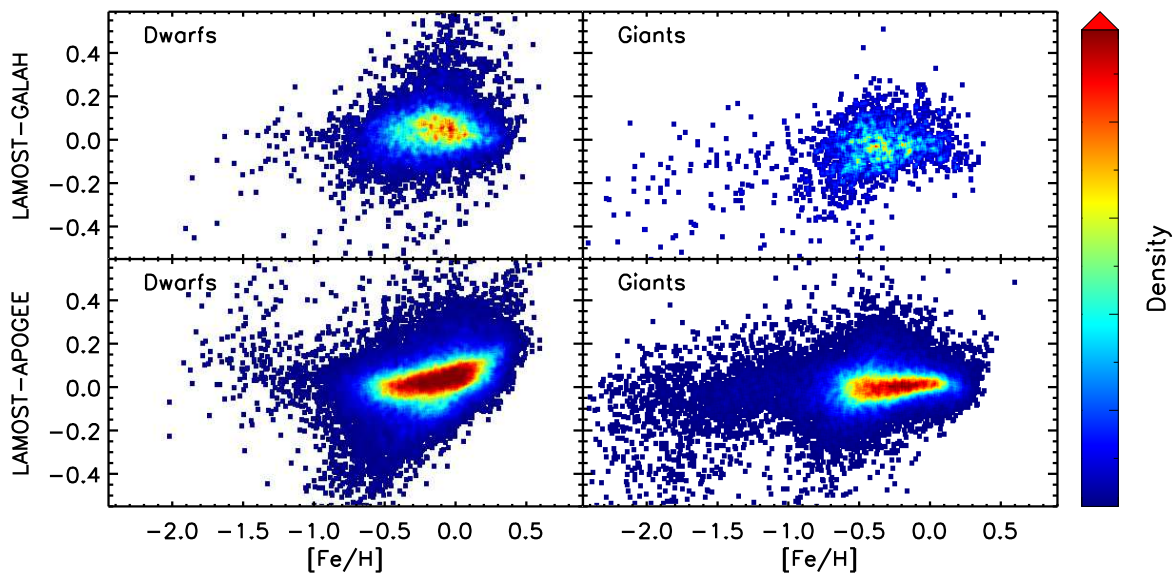}
   \caption{ [Fe/H] metallicity difference of common stars as a function of LAMOST metallicity. Giants and dwarfs are separated by adopting the criteria of logg < 3 and logg > 3, respectively. }
   \label{star_feh}
\end{figure*}

\subsubsection{ Verifying metallicities }
We compared the [Fe/H] metallicity between our catalog and DJ20. In Figure~\ref{cmp_feh}, there are 38 common clusters whose [Fe/H] uncertainty in our catalog are not zero and we find a mean offset in [Fe/H] of -0.02 dex and a scatter of 0.10 dex. We note that all discrepant values are come from clusters with the lower number of stars for estimation. Excluding clusters whose number of stars for estimation are less than 10, our result shows good agreement with APOGEE result.

Furthermore, we note that the offset shows a tiny gradient along the metallicity in Figure~\ref{cmp_feh}. In order to study the origin of this trend, we compare the metallicity difference of common stars between LAMOST DR5 and other spectroscopic catalogs (GALAH DR2 and APOGEE DR16). To reduce the effect of stars with low SNR, we only select common stars whose LAMOST SNR are greater than 10 for comparison. Table~\ref{feh_offset} list the comparison results of metallicity offset and dispersion. The overall small offsets and dispersion indicate the reliability of metallicity measurement in LAMOST DR5 since they are in good agreement with high resolution spectroscopic results. 

In Figure~\ref{star_feh}, we plot the stellar [Fe/H] metallicity difference between LAMOST DR5 and GALAH DR2 and APOGEE DR16. We note that the [Fe/H] difference of dwarfs between LAMOST and APOGEE shows positive gradient along the metallicity, which also indicate the trend in Figure~\ref{cmp_feh} may come from the measurement difference of dwarfs between the two catalogs.

\begin{figure}
   \centering
   \includegraphics[angle=0,scale=1.2]{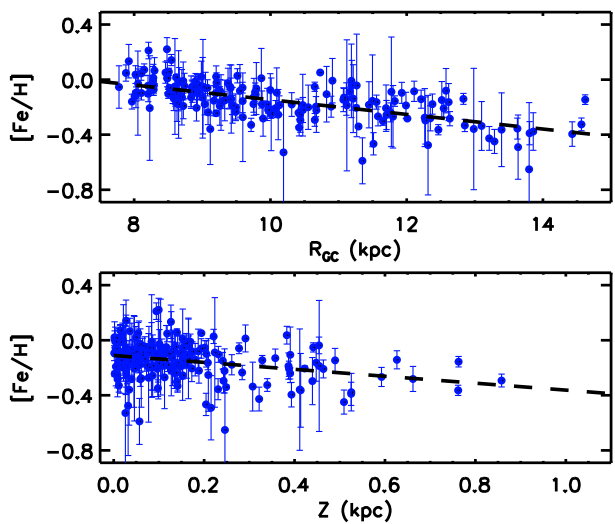}
   \caption{ Radial (upper panel) and vertical (bottom panel) metallicity gradient of young open clusters. The slope of gradients are -0.053 $\pm$ 0.004 dex kpc$^{-1}$ and -0.252$\pm$ 0.039 dex kpc$^{-1}$, respectively.}
   \label{gradient}
\end{figure}

\begin{figure}
   \centering
   \includegraphics[angle=0,scale=1.15]{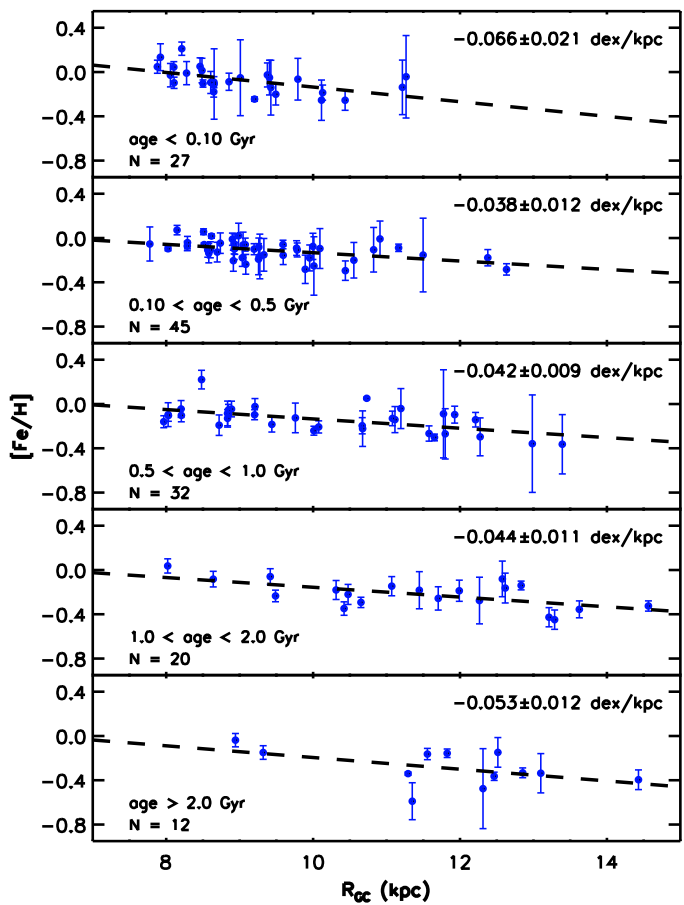}
   \caption{Radial metallicity gradients in different age bins. Dashed lines are linear least-squares approximation in one-dimension with [Fe/H] errors. }
   \label{radial_time}
\end{figure}

\begin{figure}
   \centering
   \includegraphics[angle=0,scale=1.15]{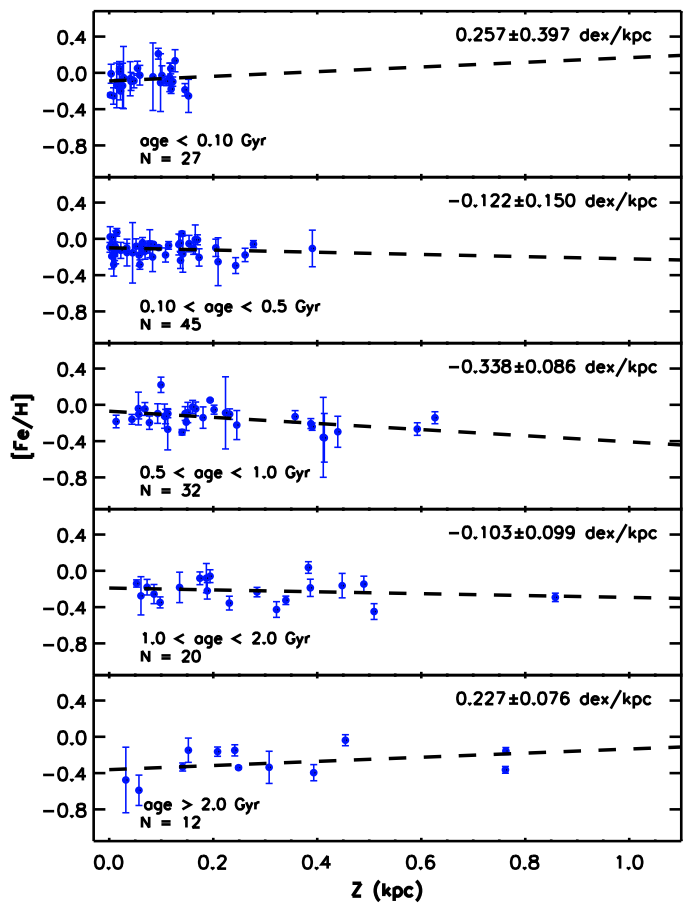}
   \caption{ Vertical metallicity gradients in different age bins. Symbols are the same as in Figure~\ref{radial_time}}
   \label{vertical_time}
\end{figure}

\begin{figure}
   \centering
   \includegraphics[angle=0,scale=1.1]{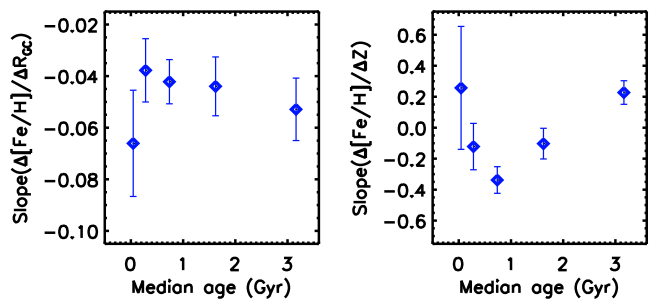}
   \caption{Radial (left panel) and vertical (right panel) metallicity gradient trends along the median age of each age bin.}   
   \label{gradient_sum}
\end{figure}

\begin{figure}
   \centering
   \includegraphics[angle=0,scale=1.1]{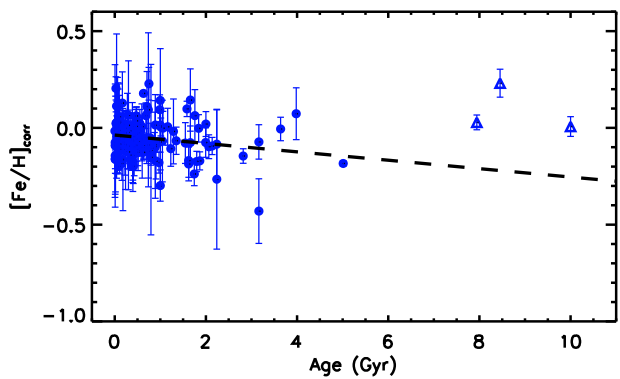}
   \caption{Age-metallicity relation of open clusters. In our sample, the slope for open clusters with age < 6 Gyr is -0.022 $\pm$ 0.008 dex Gyr$^{-1}$. Three outliers are marked as triangles and excluded from the linear fitting procedure.}
   \label{amr}
\end{figure}

\section{Abundance analysis}
\label{p}

\begin{table}
\caption{Summary of reported radial metallicity gradients using open clusters as tracers}
\label{ref}
 \centering
\begin{tabular}{cccc}
\hline
 Slope & Range & Number & ref.\\
 dex kpc$^{-1}$ & kpc &  &  \\
\hline
-0.053 $\pm$ 0.004  &  7-15 & 183 & this work \\
-0.061 $\pm$ 0.004 & 7-12 & 18 & \citet{2018AJ....156..142D} \\
-0.052 $\pm$ 0.011  &  < 12 & 79  & \citet{2016MNRAS.463.4366R} \\
-0.056 $\pm$ 0.007  &  < 17 & 488   & \citet{2009MNRAS.399.2146W} \\
-0.063 $\pm$ 0.008  &  < 17 & 118 & \citet{2003AJ....125.1397C} \\
-0.059  $\pm$ 0.010   &  7-16 & 39   & \citet{2002AJ....124.2693F} \\
-0.085  $\pm$  0.008 & 7-16 & 37 & \citet{1998MNRAS.296.1045C} \\
\hline
\end{tabular}
\end{table}

\begin{table}
\caption{Pearson correlation coefficients of radial and vertical metallicity gradients  in different age bins. }
\label{pcc}
 \centering
\begin{tabular}{ccc}
\hline
 Age range & Radial & Vertical \\
 Gyr & & \\
\hline
< 0.1     & -0.55  &  0.11 \\
0.1-0.5  & -0.47  &  -0.12 \\ 
0.5-1.0  & -0.56  &  -0.45 \\
1.0-2.0   & -0.61  &  -0.16 \\
> 2.0     & -0.50  &  0.34  \\
\hline
\end{tabular}
\end{table}

\subsection{Radial metallicity gradient}
Radial metallicity gradient in the Galactic disk plays an important role in studying the chemical formation and evolution of the Galaxy. In addition of stars or planetary nebulae (PNe) \citep[e.g.,][]{2011AJ....142..136L,2014A&A...565A..89B}, open clusters are ideal tracers of the radial metallicity gradient study, since they have a wide span of age and distance, their coeval member stars have small metallicity dispersion. From open cluster sample in previous works, the radial metallicity gradients range from -0.052 to -0.063 dex kpc$^{-1}$ within 12 kpc \citep{2003AJ....125.1397C,2009MNRAS.399.2146W,2010A&A...511A..56P,2016MNRAS.463.4366R,2016A&A...585A.150N}. 

In our sample, most of open clusters are younger than 3 Gyr. We use these clusters to fit the average radial metallicity gradient of young component in the Galactic disk. The upper panel in Figure~\ref{gradient} shows the metallicity gradient in the Galactocentric distance range R$_{GC}$ = 7-15 kpc, with a linear fit to the whole range. Although the radial metallicity gradient of -0.053$\pm$0.004 dex kpc$^{-1}$ in the radial range 7-15 kpc is consistent with the previous works (see table~\ref{ref} for more details and comparison),  the Pearson correlation coefficient of -0.33 indicate a weak correlation for overall radial metallicity gradient of all clusters, which may be caused by the mixture of open clusters with different populations.  

In order to constraint the Galactic chemo-dynamical model, the study of gradient evolution in the Galactic disk is important \citep{1998MNRAS.296.1045C,2003AJ....125.1397C,2012AJ....144...95Y}. Figure~\ref{radial_time} shows the radial metallicity gradients in different age bins. Since we have a sufficient number of clusters in different age bins, we can perform the analysis of gradient evolution. We separate our samples into five age bins, including very young age bin (< 0.1 Gyr), from young to intermediate age bins (0.1-0.5 Gyr, 0.5-1.0 Gyr, 1.0-2.0 Gyr), and old age bin (> 2.0 Gyr). Table~\ref{pcc} show the Pearson correlation coefficient of radial metallicity gradients in different age bins. After separating clusters with age bins, the Pearson correlation coefficient show that the correlation of metallicity gradients in different age bins are stronger than the correlation of overall metallicity gradient, which also indicate the higher reliability of radial metallicity gradients in different age bins. The gradient trend with median age of each sub-sample is shown in the left panel of Figure~\ref{gradient_sum}. Ignoring very young sample, the rest of four age samples display a mild flat trend of radial metallicity gradient with time. For clusters with age greater than 0.1 Gyr (most of them less than 4 Gyr), the steeper gradient of older population is consistent with previous studies  \citep[e.g.,][]{1998MNRAS.296.1045C,2002AJ....124.2693F,2020AJ....159..199D}. The time-flattening tendency may be explained by the common influence of radial migration  \citep{2016A&A...585A.150N,2017A&A...600A..70A} and chemical evolution in the Galactic disk \citep{2000ASSL..255..505T,2002ChJAA...2..226C, 2016A&A...591A..37J}.

However, we notice that there is a steep gradient for very young samples (< 0.1 Gyr), which is not consistent with previous results \citep{2011A&A...535A..30C,2017A&A...601A..70S} and the corresponding explanation \citep{2020A&A...634A..34B}. Although there is no convincing explanation for this reverse trend, this result is not contradictory to the chemo-dynamical simulation of \citet[MCM]{2013A&A...558A...9M,2014A&A...572A..92M}. In the MCM model, radial migration is expected to flatten the chemical gradients for ages > 1 Gyr, while also predicts an almost unchanged gradient for the very young population.  Since there is no process that has a significant impact on the gradient of very young population, their steep gradient partly represent the current chemical gradient in the Galactic disk (R$_{GC}$ $\sim$ 8-12 kpc). 

In particular, it is noteworthy that the cluster NGC6791 include in our initial sample. As many previous works noticed, this cluster is very metal-rich and fairly old \citep{1994A&A...287..761C,2014AJ....148...61T,2018AJ....156..142D}, and believed to be migrated to its current location \citep{2017ApJ...842...49L}.  In order to reduce the influence of outlier on gradients, we excluded NGC6791 from our cluster sample, and then perform the radial and vertical gradient analysis in Figure~\ref{gradient} - \ref{gradient_sum}.

\begin{table*}
\caption{Description of the open cluster properties catalog. }
\label{cat_ocs}
 \centering
\begin{threeparttable}
\begin{tabular}{llll}
\hline
Column &  Format  & Unit   & Description \\
\hline
CLUSTER & string & - & Cluster name \\
RA & float & deg & Mean right ascension of members in CG18 (J2000)\\
DEC & float & deg & Mean declination of members in CG18 (J2000)\\
PMRA & float & mas yr$^{-1}$ & Mean proper motion along RA of members in CG18 \\
PMRA\_std & float & mas yr$^{-1}$ & Standard deviation of pmRA of members in CG18 \\
PMDE & float & mas yr$^{-1}$ & Mean proper motion along DE of members in CG18 \\
PMDE\_std & float & mas yr$^{-1}$ & Standard deviation of pmDE of members in CG18 \\
DMODE & float & pc & Most likely distance of clusters in CG18 \\
RV & float & km s$^{-1}$ & Mean radial velocity measured from member spectra in LAMOST \\
RV\_std & float & km s$^{-1}$ & Standard deviation of RV \\
RV\_num & integer & - & Number of stars used for RV estimation \\
RV\_flag & String & - &  Flag of Gaussian fitting process for RV estimation\\ 
FEH & float & dex & Mean [Fe/H] measured from member spectra in LAMOST \\
FEH\_std & float & dex & Standard deviation of [Fe/H] \\
FEH\_num & integer & - & Number of stars used for [Fe/H] estimation \\
FEH\_flag & String & - &  Flag of Gaussian fitting process for [Fe/H] estimation\\ 
GX & float & pc & Galactocentric coordinate points to the direction opposite to that of the Sun \\
GX\_err & float & pc & Mean errors of GX coordinate calculation\\
GY & float & pc & Galactocentric coordinate points to the direction of Galactic rotation \\
GY\_err & float & pc & Mean errors of GY coordinate calculation\\
GZ & float & pc & Galactocentric coordinate points toward the North Galactic Pole   \\
GZ\_err & float & pc & Mean errors of GZ coordinate  calculation \\
U & float & km s$^{-1}$ & Galactocentric space velocity in X axis \\
U\_err & float &  km s$^{-1}$ & Mean errors of U velocity calculation\\
V & float & km s$^{-1}$ & Galactocentric space velocity in y axis \\
V\_err & float &  km s$^{-1}$ & Mean errors of V velocity calculation\\
W & float & km s$^{-1}$ & Galactocentric space velocity in Z axis  \\
W\_err & float &  km s$^{-1}$ & Mean errors of W velocity calculation\\
R$_{\rm ap}$ & float & pc & Averaged apogalactic distances from the Galactic centre \\
R$_{\rm peri}$ & float & pc & Averaged perigalactic distances from the Galactic centre \\
EC & float & pc & Eccentricity calculated as e=(R$_{\rm ap}$-R$_{\rm peri}$) / (R$_{\rm ap}$+R$_{\rm peri}$) \\
ZMAX & float & pc & Averaged maximum vertical distances above the Galactic plane \\
R$_{\rm gc}$ & float & pc & Galactocentric distance assuming the Sun is located at 8340 pc \\
R$_{\rm gc}$\_err & float & pc & Mean errors of Galactocentric distance calculation \\
AGE\_ref & float & Gyr &  Age from literature results determined by the isochrone fit \\
DIST\_ref & float & pc & Distance from literature results determined by the isochrone fit \\
EBV\_ref & float & - & Reddening from literature results determined by the isochrone fit \\
REF~\tnote{1}& String & - & Label of referred literature for age, distance and EBV determination \\ 
\hline
\end{tabular}
\begin{tablenotes}
       \footnotesize
       \item[1] Three labels are used to refer different literatures:
        (1)= \citet{2019A&A...623A.108B}; 
        (2)=\citet{2013A&A...558A..53K};
        (3)=\citet{2002A&A...389..871D} 
\end{tablenotes}
\end{threeparttable} 
\end{table*}

\begin{table*}
\caption{Description of the spectroscopic catalog of cluster members }.
\label{cat_mem}
 \centering
 \begin{threeparttable}
\begin{tabular}{llll}
\hline
Column &  Format  & Unit   & Description \\
\hline
OBSID & string & - & Object unique spectra ID in LAMOST DR5 \\
DESIGNATION & string & - & Object designation in LAMOST DR5 \\
RA\_obs & float & deg & Object right ascension in LAMOST DR5 (J2000) \\
DEC\_obs & float & deg & Object declination in LAMOST DR5 (J2000) \\
SNRG & float & - & Signal-to-noise ration of g filter in LAMOST spectrum \\
SNRR & float & - & Signal-to-noise ration of r filter in LAMOST spectrum \\
SNRI & float & - & Signal-to-noise ration of i filter in LAMOST spectrum \\
RV\_2d & float & km s$^{-1}$ & Radial velocity derived by the LAMOST 2D pipeline \\
RV\_2d\_err & float & km s$^{-1}$ &  Uncertainty of radial velocity derived by the LAMOST 2D pipeline \\
RV\_1d & float & km s$^{-1}$ & Radial velocity derived by the LAMOST 1D pipeline \\
RV\_1d\_err & float & km s$^{-1}$ & Uncertainty of radial velocity derived by the LAMOST 1D pipeline \\
TEFF & float & k & Effective temperature derived by the software of ULYSS\\
TEFF\_err & float & k & Error of effective temperature derived by the software of ULYSS\\
LOGG & float & dex & Surface gravity derived by the software of ULYSS \\
LOGG\_err & float & dex & Error of surface gravity derived by the software of ULYSS\\
FEH & float & dex & [Fe/H] derived by the the software of ULYSS\\
FEH\_err & float & dex & Error of [Fe/H] derived by the software of ULYSS\\
SOURCE & string & - &  Gaia DR2 source id \\
PARALLAX & float & mas & Parallax in Gaia DR2 \\  
PARALLAX\_err & float & mas & Parallax error in Gaia DR2 \\  
PMRA & float & mas yr$^{-1}$ & Proper motion along RA in Gaia DR2 \\
PMRA\_err & float & mas yr$^{-1}$ & Error of pmRA in Gaia DR2 \\
PMDE & float & mas yr$^{-1}$ & Proper motion along DE in Gaia DR2 \\
PMDE\_err & float & mas yr$^{-1}$ & Error of pmDE in Gaia DR2 \\
GMAG & float & mag & G-band magnitude in Gaia DR2 \\
BP\_RP & float & mag & BP minus RP color in Gaia DR2 \\
PROB & float & - & Membership probability provided by CG18 \\
CLUSTER & string & - & Corresponding cluster name \\
\hline
\end{tabular}
\end{threeparttable} 
\end{table*}

\subsection{Vertical metallicity gradient}

The vertical metallicity gradient is another important clue to constrain the formation history of the Galactic disk, while its existence among old open clusters was controversial \citep{1995ARA&A..33..381F,1995AJ....110.2813P}. The bottom panel in Figure~\ref{gradient} show the vertical metallicity gradient of our clusters within 1 kpc distance from the Galactic mid-plane. The resulting slope is -0.252$\pm$ 0.039 dex kpc$^{-1}$, which is in good agreement with previous results \citep[e.g,][]{1998MNRAS.296.1045C,2003AJ....125.1397C}. 

As \citet{1998MNRAS.296.1045C} pointed out, the cluster sample that they used for deriving the vertical gradient is significantly biased, because of the tidal disruption, which is more effective when closer to the Galactic mid-plane. In order to disentangle the effect of age dependence,  we plot the vertical gradients in different age bins in Figure~\ref{vertical_time}, and the gradient trend along the median age of each age sample in Figure~\ref{gradient_sum} (right panel), while age bins are the same as in radial gradient analysis. The Pearson correlation coefficients of vertical metallicity distribution with different age bins are presented in Table~\ref{pcc}, which show weak correlation or even no correlation.  It is worth noting that the vertical distribution of open clusters is effected by the different scale-heights of different age population \citep{1996A&A...310..771N}. For very young samples (< 0.1 Gyr), the positive gradient maybe caused by the small scale-height, which also leads a large dispersion of the trend. For old samples (> 2 Gyr), we suppose the positive gradient is the result of both migration and tidal disruption. Therefore, this suggests that open clusters with intermediate ages provide more reliable trend of vertical metallicity gradient than other age population.    

\subsection{Age metallicity relation}
The age-metallicity relation (AMR) is a useful clue for understanding the history of metal enrichment of the disk and providing an important constraint on the chemical evolution models. During past two decades, many works are focused on this study, either use nearby stars \citep{2001A&A...377..911F,1998MNRAS.296.1045C,1993A&A...275..101E} or use open clusters with multiple ages \citep{2016A&A...585A.150N,2003AJ....125.1397C,1998MNRAS.296.1045C}. In general, the observational data shows the evidence of decreasing metallicity with increasing age for both tracers, which indicate in principle the metal-enrichment in the interstellar medium (ISM) during the chemical evolution of the Galaxy.   

Comparing with the nearby stars, the open clusters have great advantage to identify the AMR since their metallicities and ages can be relatively more reliably determined. However, even based on the open clusters, the existence of AMR on the disk is not significant \citep{2009A&A...494...95M,1994A&A...287..761C, 1985A&A...147...47C}. For some studies, only a mild decrease of the metal content of clusters with age is found \citep{2016A&A...585A.150N,2010A&A...511A..56P,2003AJ....125.1397C}. 

Figure~\ref{amr}  shows the age-metallicity relation of  open clusters in our catalog. Ages were determined by visual inspection through the best fitting isochrone in the color-magnitude diagram (See section~\ref{age}). To remove the effect of the spatial variation of the metallicity due to the radial metallicity gradients, we build up a AMR in which we correct our [Fe/H] with the following relation [Fe/H]$_{\rm corr}$=[Fe/H]-0.053 (R$_{\odot}$-R) (kpc). After excluding 3 old open clusters as outliers, we perform the linear fitting of open clusters in our sample. The metallicity decreases with 0.022 $\pm$ 0.008 dex Gyr$^{-1}$ for open clusters within 6 Gyr. The Pearson correlation coefficients of -0.28 also indicate the weak correlation of AMR, which is consistent with the mild decrease relation in previous works \citep[e.g.,][]{2016A&A...585A.150N,2010A&A...511A..56P,2003AJ....125.1397C}.

We noted that there are three very old but metal-rich open clusters in our sample (triangles in Figure~\ref{amr}), with age in 8 Gyr or older. One of the possible explanation about the origin of these open clusters is the infalling or merger events within the time of 3-5 Gyr \citep{1998MNRAS.296.1045C}. For open clusters with age > 8 Gyr, it is suggested that they might be related by the formation of the triaxial bar structure \citep{1996A&A...310..771N} and further migrated to the current position. 

\section{Description of the catalog}
\label{cat}
We provide two catalogs\footnote{The catalogs can be download via http://dr5.lamost.org/doc/vac. Electronic versions are also available alongside the article.} in this paper: one for the properties of 295 open clusters and the other for spectroscopic parameters of 8811 member stars.    

Table~\ref{cat_ocs}  describes the catalog of open cluster properties. Columns 2-8 list astrometic parameters of open clusters provided by CG18, including the coordinates, mean proper motions, and distances, which were mainly based on the Gaia solution. Columns 9-16 list the measurement results of radial velocity and metallicity by LAMOST DR5. Columns 17-34 list derived kinematic and orbital parameters of open clusters. Columns 35-38 list parameters by the isochrone fit results in literature, including age, distance and reddening.    

Table~\ref{cat_mem} describes the spectroscopic catalog of cluster members, including the LAMOST spectra information (columns 1-7), the derived stellar fundamental parameters by the LAMOST spectra (columns 8- 17), the astrometric and photometric parameters in Gaia DR2 (columns 18-26) and the membership probability in CG18 (columns 27). 

\section{Summary}
We have used the identified cluster members by CG18 to cross-match with the LAMOST spectroscopic catalog. A total of 8811 member stars with spectrum data were provided. Using the spectral information of cluster members, we also provide average radial velocity of 295 open clusters and metallicity of 220 open cluster s. Considering the accurate observed data of tangential velocity provided by Gaia DR2 and radial velocity provided by LAMOST DR5, we further derived the 6D phase positions and orbital parameters of 295 open clusters. The kinematic results shows that most of open clusters in our catalog are located on the thin disk and have approximate circular motions. In addition,  referring to the literature results of using isochrone fitting method,  we estimated the age, distance and reddening of our sample of open clusters. 

As an value-added catalog in LAMOST DR5, the provided list of cluster members make a correlation between the LAMOST spectra and the cluster overall properties, especially for stellar age, reddening and distance module. Comparing with the  spectra of field stars, the LAMOST spectra of member stars are valuable source to perform the detail study of stellar physics or to calibrate the stellar fundamental parameters, since the cluster can provide statistical information for these members with higher precision. 

Furthermore, using the open clusters as tracers, we make use of their metallicities to study the radial metallicity gradient and the age-metallicity relation. The derived radial metallicity gradient for young clusters is -0.053$\pm$0.004 dex kpc$^{-1}$ within the radial range of 7-15 kpc, which is consistent with previous works. After excluding 3 old but metal-rich open clusters, we derived an AMR as -0.022$\pm$0.008 dex Gyr$^{-1}$ for young clusters, which follow the tendency that younger clusters have higher metallicities, as a consequence of the more enriched ISM from which they formed \citep{2009A&A...494...95M}. On the other hand, considering that the metallicity increasing of the disk is mild during the past 5 Gyr \citep{2003AJ....125.1397C}, which is indeed in agreement with our findings that a small increase in the youngest clusters, the nature of AMR of open clusters need further investigations.   

{\bf Acknowledgments}
We are very grateful to the referee for helpful suggestions, as well as the correction for some language issues, which have improved the paper significantly.
This work supported by National Key R\&D Program of China No. 2019YFA0405501. The authors acknowledges the National Natural Science Foundation of China (NSFC) under grants U1731129  (PI: Zhong), 11373054 and 11661161016 (PI: Chen) and .
Guoshoujing Telescope (the Large Sky Area Multi-Object Fiber Spectroscopic Telescope LAMOST) is a National Major Scientific Project built by the Chinese Academy of Sciences. Funding for the project has been provided by the National  Development and Reform Commission. LAMOST is operated and managed by the   National Astronomical Observatories, Chinese Academy of Sciences.
This work has made use of data from the European Space Agency (ESA) mission Gaia (\url{https://www.cosmos.esa.int/gaia}), processed by the Gaia Data Processing and Analysis Consortium (DPAC,\url{https://www.cosmos.esa.int/web/gaia/dpac/consortium}). Funding for the DPAC has been provided by national institutions, in particular the institutions participating in the Gaia Multilateral Agreement.

\end{document}